\documentclass[pra,twocolumn,tightenlines,nofootinbib,nolongbibliography]{revtex4-2}
\usepackage{bm,dcolumn,amsmath,graphicx,amsfonts,amssymb}

\usepackage{hyperref}
\usepackage{xcolor}
\usepackage{comment}
\definecolor{cite}{rgb}{0.,0.,0.9}   
\hypersetup{colorlinks,linkcolor={cite},citecolor={cite},urlcolor={cite}}

\renewcommand{\v}[1]{\ensuremath{\boldsymbol{#1}}}		
\newcommand{\abs}[1]{\ensuremath{\left |#1\right |}}
\newcommand{\bra}[1]{\ensuremath{\langle #1|}}	
\newcommand{\ket}[1]{\ensuremath{|#1\rangle}}	

\def\d{\ensuremath{{\rm d}}}

\usepackage[normalem]{ulem}
\usepackage{xspace}

\definecolor{newc}{rgb}{0.,0.6,0.4}

\begin{document}

\title{QED radiative corrections to electric dipole amplitudes in heavy atoms}

\author{C.\ J.\ Fairhall}
\author{B.\ M.\ Roberts}
\author{J.\ S.\ M.\ Ginges}
\affiliation{School of Mathematics and Physics, The University of Queensland, Brisbane QLD 4072, Australia}
\date{\today}

\begin{abstract}\noindent
We use the radiative potential method to perform a detailed study of quantum electrodynamics (QED) radiative corrections to electric dipole (E1) transition amplitudes in heavy alkali-metal atoms Rb, Cs, Fr, and alkali-metal-like ions Sr$^+$, Ba$^+$, and Ra$^+$. The validity of the method is checked by comparing with the results of rigorous QED in simple atomic potentials.
We study the effects of core relaxation, polarization of the core by the E1 field, and valence-core correlations on QED, which are shown to be important in some cases. We identify several transitions for which the QED contribution exceeds the deviation between atomic theory and experiment.
\end{abstract}

\maketitle


\section{Introduction}

In the last years, increasing attention has been given to the role of quantum electrodynamics (QED) radiative corrections in high-precision studies of heavy and superheavy many-electron atoms and ions.
They have been considered for binding and excitation energies~\cite{Pyykko2003,FlambaumQED2005,Shabaev2013,Ginges2016a,Ginges2016b} (see also calculations in multiply-charged ions~\cite{Tupitsyn2016}),
for electric dipole (E1) amplitudes~\cite{Sapirstein2005,FlambaumQED2005,Roberts2013,ShabaevPRL2005,*Shabaev2005},
and for the hyperfine structure~\cite{Sapirstein2003b,Pyykko2003,Sapirstein2006,Sapirstein2008,Ginges2017}. Account of QED radiative corrections was critical in the interpretation of the atomic parity violation measurement in cesium~\cite{Wieman1997,Kuchiev2002,Milstein2002} (see also Ref.~\cite{GingesCs2002,ShabaevPRL2005,*Shabaev2005,FlambaumQED2005,Porsev2009,*Porsev2010,DzubaCsPNC2012,Roberts2022}), and was important for resolving discrepancies between theory and experiment for the ionization potential and electron affinity for gold~\cite{Pasteka2017}. See Ref.~\cite{Schwerdtfeger2015} for a review on QED corrections to superheavy elements.

The interplay between QED and many-body effects in heavy many-electron systems has been explored for the binding and excitation energies~\cite{Pyykko2003,FlambaumQED2005,Ginges2016a,Ginges2016b,Shabaev2013}. No such detailed study of combined QED and many-body effects has been performed for the E1 amplitudes to date, while it is known~\cite{FlambaumQED2005,Sapirstein2005,Roberts2013} that such corrections are significant at the level of accuracy of state-of-the-art calculations (several 0.1\%).
While an approximate ``radiative potential" has been used to evaluate QED corrections to E1 amplitudes (see, e.g., Refs.~\cite{FlambaumQED2005,Roberts2013,Sahoo2021,Roberts2022}),
the validity of such an approach has never been carefully
checked against rigorous quantum electrodynamics.
The ability to theoretically describe E1 amplitudes with high precision is important in a number of different areas, both fundamental and applied, including in studies of violations of fundamental symmetries and searches for new physics ~\cite{GingesRev2004,RobertsReview2015,AtomicReview2018}, and atomic polarizabilities and application to atomic clocks~\cite{MitroyReview2010,DereviankoReview2011,LudlowReview2015} (see also Ref.~\cite{Safronova2014AtomicClocks}).

The goals of the current paper are as follows:\ first, to check the validity of the radiative potential approach for evaluating QED radiative corrections to E1 amplitudes in heavy atoms by comparing with the results of rigorous QED using simple atomic potentials.
Second, to study in detail the combined QED and many-body effects to better understand the mechanisms and identify which effects are most important and should be taken into account in accurate studies.
Third, to identify those amplitudes where the relative QED corrections are particularly large.
And finally, to assess the importance of account of QED corrections at the level of atomic theory precision.
We find that in several cases the QED contribution exceeds the deviation between atomic theory and experiment. That is, atomic theory has reached the level of precision where the QED radiative corrections to E1 amplitudes may be observed in experiments with heavy atoms.

The paper is organized as follows.
In Section~\ref{sec:radPot} we describe the radiative potential method and use it to find QED corrections to E1 amplitudes for alkali-metal atoms, and we compare the results with those of rigorous QED.
In Section~\ref{sec:theory} we describe the many-body methods we use in our calculations. Our results of combined QED and many-body theory are given in Section~\ref{sec:results} for the atoms and ions Rb, Sr$^+$, Cs, Ba$^+$, Fr, Ra$^+$, and in Section~\ref{sec:exptComp} we present our results of total E1 amplitudes and compare these with measured values.

\section{The radiative potential method and comparison with rigorous QED}
\label{sec:radPot}

The radiative potential approach \cite{FlambaumQED2005,Ginges2016a,Ginges2016b} is used widely in calculations of QED radiative corrections to energies in many-electron atoms, ions, molecules; see, e.g.,  Refs.~\cite{Roberts2013,Ginges2016b,Dzuba2013,Safronova2014,Safronova2014AtomicClocks,Ginges2015,Li2021,Sunaga2022}. In essence, this approach is based on the use of a radiative potential which may be added in a simple manner to atomic many-body methods and computer codes. The radiative potential is comprised of a self-energy (SE) and vacuum polarization (VP) contribution~\cite{FlambaumQED2005},
\begin{equation}
    V_{\text{rad}}(\boldsymbol{r}) = V_{\text{SE}}(\boldsymbol{r}) + V_{\text{VP}}(r) \, .
\end{equation}
The vacuum polarization correction is well-approximated by the Uehling potential, and the self-energy correction is approximated by a local potential,
\begin{equation}
    V_{\text{SE}}(\boldsymbol{r}) = V_{\text{mag}}(\boldsymbol{r})+V_{\text{high}}(r)+V_{\text{low}}(r) \, ,
    \label{eqn:qedSE}
\end{equation}
consisting of three parts:\ magnetic, $V_{\rm mag}$, and high-frequency and low-frequency electric, $V_{\rm high}$ and $V_{\rm low}$.
The self-energy part of the radiative potential is defined such that its expectation value for hydrogen-like ions is equal to the one-loop self-energy radiative corrections to the energies for these ions. The electric parts of the potential contain fitting factors to ensure that the results of rigorous QED are reproduced. We refer the reader to Refs.~\cite{FlambaumQED2005,Ginges2016a,Ginges2016b} for details and the explicit expressions.
We use the finite-nuclear-size formulae presented in Refs.~\cite{Ginges2016a,Ginges2016b}.

The atoms and ions considered in this work have one valence electron above closed shells. We consider a number of different approximation schemes for account of the electron-electron interactions. All calculations begin by solving the Dirac equation, $h_D\varphi = \varepsilon\varphi$,
where
\begin{equation}
    h_D = c\boldsymbol{\alpha} \cdot \boldsymbol{p} + \left(\beta-1\right)c^2 + V_{\text{nuc}} + V_{\text{el}}
    \label{eqn:spHam}
\end{equation}
is the single-particle Hamiltonian, $c$ is the speed of light, $\beta$ is a $4 \times 4$ Dirac matrix, $\boldsymbol{\alpha}$ is a vector of Dirac matrices, $\boldsymbol{p}$ is the momentum operator, $\varepsilon$ is the electron binding energy, and $V_{\text{nuc}}$ and $V_{\text{el}}$ are the nuclear and electronic potentials, respectively. Atomic units $\hbar = m_e = |e| = c\alpha = 1$ are used throughout unless otherwise specified. The nuclear potential $V_{\text{nuc}}$ is found by modelling the nuclear charge with a Fermi distribution with the usual parameters, $2.3\,{\rm fm}$ for the 90\% to 10\% fall-off thickness and the root-mean-square charge radii from the compilation~\cite{Angeli2013}.

A relativistic single-electron orbital with principal quantum number $n$ and relativistic angular momentum quantum number $\kappa=(l-j)(2j+1)$, where $j$ is the total angular momentum and $m$ its projection, and $l$ is the orbital angular momentum, may be written as
\begin{equation}
    \varphi_{n\kappa m}(\v{r}) = \frac{1}{r}
    \begin{pmatrix}
        f(r)\,\Omega _{\kappa m} \\
        i\alpha g(r) \, \Omega_{-\kappa m}
    \end{pmatrix}
    ,
    \label{eqn:dirOrb}
\end{equation}
\noindent
where $f(r)$ and $g(r)$ are upper and lower radial components of the wave function, and $\Omega_{\kappa m}$ is a spherical spinor.
The electric dipole operator in the length gauge is given by
\begin{equation}
    \v{d}=-\v{r},
\end{equation}
\noindent
where $\v{r}$ is the position operator. The E1 matrix element ($z$-component) for a transition between states $a$ and $b$ may be expressed, via the Wigner-Eckart theorem, as
\begin{equation}
    \bra{\xi_a} d_z \ket{\xi_b} = \left(-1\right)^{j_a-m_a}
    \begin{pmatrix}
        j_a & 1 & j_b \\
        -m_a & 0 & m_b
    \end{pmatrix}
    \left\langle n_a \kappa_a\| d\| n_b \kappa_b\right\rangle \, ,
    \label{eqn:me}
\end{equation}
\noindent
where $\xi$ contains the quantum numbers $n$, $\kappa$ and $m$, and $\left\langle n_b \kappa_b\| d\| n_a \kappa_a\right\rangle$ is a reduced matrix element, which does not depend on projections. 
Throughout this work, we will simplify the E1 reduced matrix element notation to
\begin{equation}
    z_{ab} = \left\langle n_a \kappa_a\| d\| n_b \kappa_b\right\rangle .
\end{equation}
\noindent
The reduced E1 matrix element is composed of an angular part, $C_{ab}$, and a radial part, $r_{ab}$,
\begin{equation}
    z_{ab} = -C_{ab}r_{ab}
        =-C_{ab} \int_0^{\infty}\d r\, \left(f_a f_b + \alpha^2 g_a g_b\right)r \, .
\end{equation}
This expression corresponds to the lowest-order E1 amplitude.

By adding the radiative potential to the atomic Hamiltonian, Eq.~(\ref{eqn:spHam}), QED radiative corrections to the energies and to the wave functions are obtained. These wave functions, perturbed by the QED interactions, may then be used in the matrix elements to yield the QED corrections to the E1 amplitudes in the radiative potential approach.
Throughout this work, the radiative corrections are found by performing the calculations with and without the radiative potential, and taking the difference in the results,
\begin{equation}
\label{eq:QEDcorr}
\delta_{\rm QED}= \left\langle \widetilde{n_a \kappa_a}|| d\| \widetilde{n_b \kappa_b}\right\rangle - \left\langle n_a \kappa_a\| d\| n_b \kappa_b\right\rangle \, ;
\end{equation}
the wave functions perturbed by the radiative potential are indicated by the tilde.
This is how the QED corrections are found at any considered level of many-body theory.

\begin{figure}
    \includegraphics[width=0.65\columnwidth]{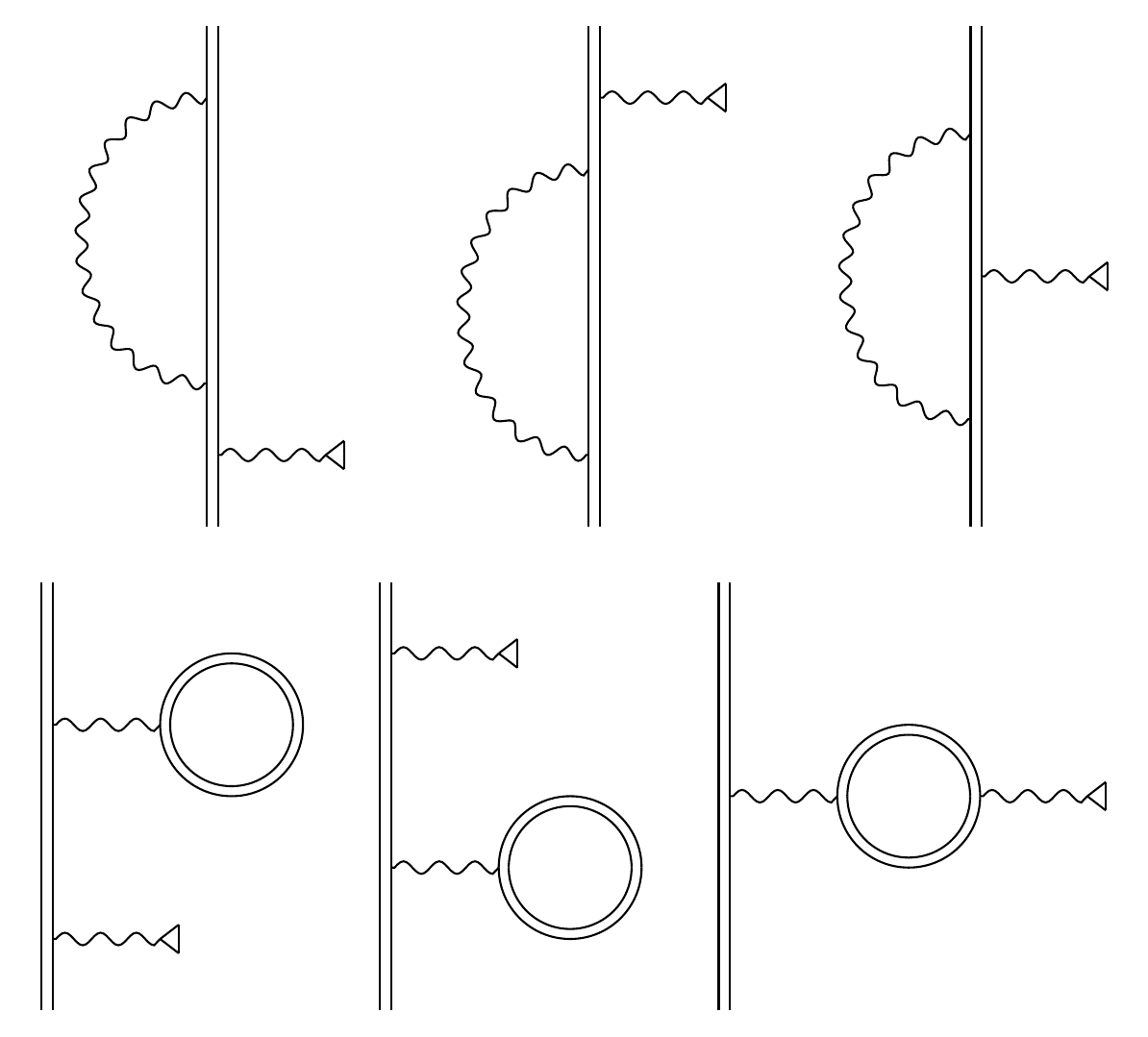}
    \caption{Feynman diagrams for QED corrections to E1 amplitudes. Top row: self-energy corrections; bottom row: vacuum polarization corrections. Wavy lines with triangles represent the E1 field, wavy lines the photon propagator, and double line the bound electron wave function and propagator. Diagrams on the right correspond to vertex corrections.}
    \label{fig:feynQED}
\end{figure}

In Fig.~\ref{fig:feynQED}, the one-loop self-energy and vacuum polarization corrections to the E1 amplitude are illustrated. The left and middle diagrams give the SE and VP contributions that correspond to corrections to the electron wave functions. The diagrams on the right are the vertex corrections, and cannot be included through corrections to wave functions. From the ``low-energy theorem'', the vertex corrections are expected to be small compared to the ``perturbed orbital'' corrections (see, e.g., Ref.~\cite{FlambaumQED2005} for discussion). Indeed, in this case, the electric dipole operator, which acts at large distances, is essentially ``locked'' inside the photon loop located at small distances from the nucleus. Therefore, it is the perturbed-orbital corrections -- those that may be found using a radiative potential -- that are anticipated to give the largest contributions.

To gauge the accuracy of the radiative potential approach for evaluation of QED radiative corrections to E1 amplitudes, we calculate the corrections in the same  atomic potential used in rigorous QED calculations and compare the results.
The only such rigorous QED calculations for many-electron atoms were performed by Sapirstein and Cheng~\cite{Sapirstein2005};
the calculations were performed for alkali-metal atoms in the Kohn-Sham atomic potential.
In the Kohn-Sham approximation, the electronic potential $V_{\text{el}}$ in Eq.~(\ref{eqn:spHam}) is approximated by
\begin{equation}
    V_{\text{el}}(r) = \int  \d r^{\prime}\ \frac{\rho_t\left(r^{\prime}\right)}{\left|r-r^{\prime}\right|} + \dfrac{2}{3}\left[ \frac{81}{32\pi^2}r\rho_t(r)\right],
\end{equation}
\noindent
where
\begin{equation}
    \rho_t(r) = \sum_{i=1}^{N} \left(f_i^2(r) + \alpha^2 g_i^2(r)\right)
    \label{eqn:KSrho}
\end{equation}
is the total probability density formed from all $N$ electrons in the atom. In this approximation, all electrons are included in the self-consistent solution for the atomic potential. At large distances, the valence electron should ``see" the potential of a single positive charge. Since Eq.~(\ref{eqn:KSrho}) includes the valence electron, the Latter correction~\cite{Latter1955} is included explicitly to force the potential to decay as $1/r$.

\begin{table}[t]
    \caption{Self-energy corrections to E1 matrix elements between lowest $s$ and $p_{1/2}$ states of alkali-metal atoms in the Kohn-Sham potential, expressed as relative corrections $R_{ab}$; see Eq.~(\ref{eq:Rab}). Lowest-order radial integral $r_{ab}$ is given in the first column of results. Radiative potential results are separated into $s$-wave and $p$-wave perturbed-orbital contributions, ${\rm PO}(s)$ and ${\rm PO}(p)$. Results of rigorous QED calculations~\cite{Sapirstein2005} are presented for comparison in the final three columns.}
    \label{tab:qedPO}
    \begin{ruledtabular}
        \begin{tabular}{*{7}c}
            Atom & $|r_{ab}|$ & \multicolumn{2}{c}{This work} & \multicolumn{3}{c}{Sapirstein and Cheng~\cite{Sapirstein2005}} \\
            \cline{3-4} \cline{5-7}
            && PO($s$) & PO($p$) & PO($s$) & PO($p$) & vertex\footnotemark[1] \\
            \hline
            Na & 4.588 & 0.032 & 0.000 & 0.031 & 0.001 & -0.015 \\
            K & 5.681 & 0.069 & 0.000 & 0.067 & 0.000 & -0.003 \\
            Rb & 6.009 & 0.190 & 0.000 & 0.182 & 0.000 & 0.028 \\
            Cs & 6.585 & 0.334 & -0.001 & 0.326 & 0.000 & -0.065 \\
            Fr & 6.511 & 0.777 & -0.014 & 0.787 & 0.202 & -0.060 \\
        \end{tabular}
    \end{ruledtabular}
    \footnotetext[1]{Vertex and other corrections. A full breakdown of these contributions is given in Table II of Ref.~\cite{Sapirstein2005}.}
\end{table}

In Table~\ref{tab:qedPO}, we present our results for the self-energy corrections to E1 amplitudes obtained using the radiative potential alongside rigorous QED results of Sapirstein and Cheng~\cite{Sapirstein2005}. Results from sodium to francium are shown.
Following Ref.~\cite{Sapirstein2005}, the values are expressed as a relative correction $R_{ab}$,
\begin{equation}
\label{eq:Rab}
    r_{ab} + \delta r_{ab}  = r_{ab}\left(1 + \frac{\alpha}{\pi}R_{ab}\right) \, .
\end{equation}
It is seen from both works that for the considered systems the perturbed-orbital contribution is given in its entirety by the QED-perturbed $s$ orbitals, except for the case of francium. For the $s$-perturbed contributions, the results of our work are in excellent agreement with those of Ref.~\cite{Sapirstein2005}, differing by only $\sim 1\%$. For Fr, the $p$-perturbed contribution is different in magnitude and sign, and according to Ref.~\cite{Sapirstein2005} contributes as much as 20\% to the total SE correction, while in our evaluation it is much smaller; the reason for this difference is not clear.
It is seen from the results of Ref.~\cite{Sapirstein2005} that account of the vertex contribution is important for Na though less so for the heavier elements, where it contributes from just a few percent of the total SE correction for some elements and up to 25\% for Cs. From this comparison, we then estimate that the uncertainty associated with the use of the radiative potential for evaluation of QED corrections to E1 amplitudes for atoms from potassium and heavier is about 25\%. In the following sections we consider the effect of account of many-body effects on the QED correction.

\section{Many-body methods}
\label{sec:theory}

For heavy atoms, the QED contribution to E1 matrix elements comes mostly from perturbed-orbital corrections to $s$-states, which the radiative potential method accurately models; see Table~\ref{tab:qedPO}.
As we show in this section, the combination of many-body effects with QED is often more important than the missed effects.
The most important is the so-called relaxation effect, which is dominated by perturbed-orbital corrections to $s$-states in the core.
Such effects would be cumbersome to evaluate in rigorous QED, though straightforward using the radiative potential approach.

In this section we outline the many-body effects that are taken into account alongside the QED corrections in this work.
 In Sections~\ref{ssec:rhfCoreRel},\,\ref{ssec:corePol},\,\ref{ssec:correlations} we describe the effects which are included with the radiative potential (and without) to yield a QED radiative correction in the considered approximation (see Eq.~(\ref{eq:QEDcorr})) -- core relaxation, core polarization, and (fitted) second-order correlations. In Sections~\ref{ssec:correlations} and \ref{ssec:otherMB} we describe higher-order and other many-body corrections -- all-orders correlation corrections, structural radiation and normalization of states, and Breit -- that are included for a high-precision evaluation of the full E1 matrix element, the results of which we present in Section~\ref{sec:exptComp}.

\subsection{RHF and core relaxation}
\label{ssec:rhfCoreRel}

For alkali-metal atoms and alkali-metal-like ions, the starting approximation for the many-body consideration is the relativistic Hartree-Fock (RHF) approximation in the potential formed from the $N-1$ core electrons, with \begin{equation}
    V_{\text{el}}=V_{\text{HF}}^{N-1}
\end{equation}
in Eq.~(\ref{eqn:spHam}); see, e.g. Ref.~\cite{JohnsonBook2007} for explicit expressions.
In the previous section, QED radiative corrections were considered to arise due to interactions involving just the valence states. However, radiative corrections are also produced due to interactions taking place in the electron core. This is readily taken into account by adding the radiative potential to the Hartree-Fock Hamiltonian for the core,
\begin{equation}
\left(h_D+V_{\text{rad}}\right)\varphi_c^\prime=\varepsilon_c^\prime\varphi_c^\prime \,,
\end{equation}
and finding new core states $\left(\varphi_c^\prime\right)$, energies $\left(\varepsilon_c^\prime\right)$, and a new RHF potential $\widetilde{V}_{\text{HF}}^{N-1}$ from the self-consistent solutions.
New valence electron states $\varphi^\prime$ and energies $\epsilon^\prime$ may then be found in this new potential. These corrections are referred to as {\it core relaxation} radiative corrections (see Refs.~\cite{Derevianko2004,Ginges2016a,Ginges2016b} for more detail).

\subsection{Core polarization}
\label{ssec:corePol}

When subjected to an external field of frequency $\omega$, e.g.,\ an electric dipole field, the core electrons in an atom become perturbed,
\begin{equation}\label{eq:dphi}
   \varphi + \delta\varphi=
    \varphi + Xe^{-i\omega t}+Ye^{i\omega t},
\end{equation}
and a correction to the Hartree-Fock potential is produced, in a similar manner to the core relaxation mechanism described in the previous section.
The perturbation is included into calculations via the time-dependent Hartree-Fock (TDHF) method, which is equivalent to the random phase approximation with exchange (RPA)~\cite{Dzuba1984,Johnson1980,Johnson1989}.
The TDHF equations
\begin{align}
    \left(h_D - \varepsilon_c-\omega\right)X_c = -\left({d_z} + \delta V_{\text{E1}}\right)\varphi_c\\
    \left(h_D - \varepsilon_c+\omega\right)Y_c = -\left({d_z} + \delta V^\dag_{\text{E1}}\right)\varphi_c
\end{align}
are solved self-consistently for all core states;
$\delta V_{\text{E1}}$ is the correction to the RHF potential arising from E1 corrections to core orbitals.
The perturbation of the electron core by the external field is called {\it core polarization} and is an important many-body effect. The account of core polarization leads effectively to an addition to the electric dipole operator,
\begin{equation}
    \bra{a}d_z\ket{b} \rightarrow \bra{a}d_z+\delta V_{\text{E1}}\ket{b} \, .
\end{equation}
To account for the QED radiative corrections, $V_{\rm rad}$ is added to $h_D$ in the RHF and TDHF equations.

\subsection{Valence-core electron correlations}
\label{ssec:correlations}

We take into account valence-core electron correlations using the correlation potential method \cite{Dzuba1984a}.
In this method, a non-local, energy-dependent correlation potential $\Sigma({\bf r},{\bf r}^\prime,\epsilon)$ is added to the RHF equations, yielding Brueckner orbitals $\varphi^{\text{(Br)}}$ and energies $\varepsilon^{\text{(Br)}}$ for the valence electron:
\begin{equation}
\label{eqn:Brueckner}
    \left(h_D+\Sigma\right)\varphi^{\text{(Br)}}=\varepsilon^{\text{(Br)}}\varphi^{\text{(Br)}} \, .
\end{equation}
Calculation of E1 matrix elements with valence-core correlations and core polarization included corresponds to evaluation of
\begin{equation}
    \bra{\varphi_a^{\text{(Br)}}}d_z + \delta V_{\text{E1}}\ket{\varphi_b^{\text{(Br)}}} \, .
\end{equation}
A simple and effective way to account for missed higher-order many-body corrections in the correlation potential may be implemented through the introduction of a fitting factor $f$, $\Sigma \rightarrow f\Sigma$, which is found from  Eq.~(\ref{eqn:Brueckner}) by varying $f$ until the experimental binding energies are reproduced.
For evaluation of QED radiative corrections with core polarization and valence-core correlations included, we consider the correlation potential in the (lowest) second order in the Coulomb interaction, which we denote by $\Sigma^{(2)}$, with fitting included.

For the full E1 matrix element evaluation, which we perform in Section~\ref{sec:exptComp}, the all-orders correlation potential $\Sigma^{(\infty)}$ is used, and fitting is also included. The QED corrections do not change in any significant way whether the core-valence correlation potential is taken to be $\Sigma^{(2)}$ or $\Sigma^{(\infty)}$. However, inclusion of the all-orders correlation potential for the full E1 matrix elements significantly improves the theory result; with fitting included in both approaches, however, the difference between the results is very much reduced.
For evaluation of the all-orders correlation potential, we use the Feynman diagram approach.
Three classes of diagrams are included to all-orders in the Coulomb interaction: screening of the Coulomb interaction by the core electrons \cite{DzubaCPM1988pla}; the hole-particle interaction inside hole-particle loops  \cite{DzubaCPM1989plaEn}; and chaining of the correlation potential  \cite{Dzuba1984a,Dzuba1985}.
For more detail about the all-orders correlation potential method, we refer the reader to the references above or to the review~\cite{GingesRev2004}.

\subsection{Other many-body corrections to E1 amplitudes}
\label{ssec:otherMB}

In our high-precision calculations of the full E1 amplitudes, we include several other many-body corrections that enter at the level of a fraction of a percent. One of these is the Breit correction, which accounts for magnetic and retardation corrections to the electron-electron Coulomb interaction. The Breit interaction between electrons $i$ and $j$ is given by
\begin{equation}
    h_{ij}^{\text{B}}=-\frac{1}{2r_{ij}} \left(\v{\alpha}_i \cdot \v{\alpha}_j + \frac{\left(\v{\alpha}_i \cdot \v{r}_{ij}\right)\left(\v{\alpha}_j \cdot \v{r}_{ij}\right)}{r_{ij}^2}\right)\, ,
\end{equation}
\noindent
where $\v{r}_{ij}=\v{r}_i-\v{r}_j$. The Breit interaction is included into the RHF and TDHF equations, and the self-consistent solutions are used to form the second-order correlation potential; see, e.g., Ref.~\cite{Derevianko2002}.

A further many-body correction arises from diagrams corresponding to an external field (E1 field) acting on internal lines of the second-order correlation potential, termed {\it structural radiation}. At the same (third-order) level of perturbation theory, there is another correction that arises due to the change in the normalization of states. We include both structural radiation and normalization of states corrections in our calculations of all-orders E1 amplitudes; we refer the reader to Refs.~\cite{Dzuba1987jpbRPA,Blundell1987} for more details.

\section{QED results}
\label{sec:results}

We present our results for QED corrections to E1 amplitudes in three tables, highlighting the contributions of different many-body effects: first, we show the effect of account of core relaxation, which is demonstrated for $s-p$ and $p-d$ transitions for francium (Table~\ref{tab:FrQED}); then we show a detailed account of the many-body contributions to QED -- core relaxation, core polarization, second-order valence-core correlations, and fitted second-order correlations -- for cesium (Table~\ref{tab:CsQED}); and finally, we present our final QED results for the other systems considered in this work, Rb, Sr$^+$, Ba$^+$, Fr, and Ra$^+$ (Table~\ref{tab:allQED}).

\begin{table}[t]
    \caption{Vacuum polarization ($\delta_{\text{VP}}$), self-energy ($\delta_{\text{SE}}$), and total QED ($\delta_{\text{QED}}$) corrections to E1 matrix elements in Fr. Results in the frozen RHF potential (RHF$_0$) and with core relaxation included (RHF) are given. Corrections are to the absolute values of the lowest-order E1 reduced matrix elements, $\abs{z_{ab}}$, in the RHF approximation. Units: $\abs{e}a_0$.}
    \label{tab:FrQED}
    \begin{ruledtabular}
        \begin{tabular}{*{9}c}
        & & & \multicolumn{3}{c}{RHF$_0$ $\left(10^{-3}\right)$} & \multicolumn{3}{c}{RHF $\left(10^{-3}\right)$} \\
        \cline{4-6} \cline{7-9}
        $a$ & $b$ & $\abs{z_{ab}}$ & $\delta_{\text{VP}}$ & $\delta_{\text{SE}}$ & $\delta_{\text{QED}}$ & $\delta_{\text{VP}}$ & $\delta_{\text{SE}}$ & $\delta_{\text{QED}}$ \\
        \hline
        $7s_{1/2}$ & $7p_{1/2}$ & 5.144 & -2.07 & 9.04 & 7.00 & -2.28 & 9.06 & 6.82 \\
        & $7p_{3/2}$ & 7.090 & -3.40 & 14.38 & 11.04 & -3.82 & 14.70 & 10.94 \\
        $6d_{3/2}$ & $7p_{1/2}$ & 9.222 & -0.19 & 0.67 & 0.48 & 0.40 & -3.54 & -3.14 \\
        & $7p_{3/2}$ & 4.283 & 0.00 & 0.22 & 0.22 & 0.35 & -2.06 & -1.72 \\
        $6d_{5/2}$ & $7p_{3/2}$ & 12.804 & -0.01 & 1.27 & 1.26 & 0.77 & -5.09 & -4.32 \\
        \end{tabular}
    \end{ruledtabular}
\end{table}

In Table~\ref{tab:FrQED} we show the effect of core relaxation for $s-p$ and $p-d$ transitions of Fr. The calculations are performed at the RHF level of approximation. In the first instance, no QED corrections are included in the core; this is denoted by RHF$_{0}$ in the table. In the second case, the radiative potential is added to the Hamiltonian from the beginning, that is, including for the core states; we denote this by RHF in the table.
We present the results for vacuum polarization and self-energy separately and together.
As is to be expected from consideration of the QED radiative corrections to the energies~\cite{FlambaumQED2005,Ginges2016a,Ginges2016b}, the self-energy corrections dominate, and the vacuum polarization corrections are significantly smaller and of opposite sign.
The effect of core relaxation on the $s-p$ transitions is relatively small.
However, the core relaxation correction is very important for the $p-d$ transitions, changing the sign of the QED correction and increasing its size by up to an order of magnitude. It is worth noting that the core relaxation effect is seen in most cases to be larger for the (delta-function-like) Uehling potential (vacuum polarization) than for the longer-range self-energy.
The importance of accounting for core relaxation in the QED corrections to the energies is well-established; see e.g., Refs.~\cite{Derevianko2004,FlambaumQED2005,Ginges2016a,Ginges2016b}.
The core relaxation corrections arise mostly due to QED-perturbation of the $s$-states within the atomic core, influencing the non-$s$ states through the Coulomb interaction. The radiative potential accurately accounts for these corrections arising from QED-perturbation of the $s$-orbitals, as we saw in Section~\ref{sec:radPot}, Table~\ref{tab:qedPO}.

\begin{table}[t]
    \footnotesize
    \caption{Absolute QED radiative corrections $\delta_{\text{QED}}$ to E1 amplitudes for Cs in different approximations. Relativistic Hartree-Fock in the frozen potential ``RHF$_0$'', with core relaxation included ``RHF'', with core polarization ``$+\delta V_{\rm E1}$'', with second-order correlation potential ``$+\Sigma^{(2)}$'', and with fitted second-order correlation potential ``$+\lambda\Sigma^{(2)}$''. The absolute values of the E1 reduced matrix elements $\abs{z_{ab}}$ are included for reference, in both RHF and $\lambda\Sigma^{(2)}$ approximations. Units: $\abs{e}a_0$.}
    \label{tab:CsQED}
    \begin{ruledtabular}
        \begin{tabular}{*{9}c}
         & & \multicolumn{2}{c}{$\abs{z_{ab}}$} & \multicolumn{5}{c}{$\delta_{\text{QED}} \left(10^{-3}\right)$} \\
        \cline{3-4} \cline{5-9}
        \multicolumn{1}{c}{$a$} & \multicolumn{1}{c}{$b$} & \multicolumn{1}{c}{RHF} & \multicolumn{1}{c}{$\lambda\Sigma^{(2)}$} & \multicolumn{1}{c}{RHF$_0$} & \multicolumn{1}{c}{RHF} & \multicolumn{1}{c}{$+\delta V_{\rm E1}$} & \multicolumn{1}{c}{$+\Sigma^{(2)}$} & \multicolumn{1}{c}{$+\lambda\Sigma^{(2)}$} \\

        \hline
        $6s_{1/2}$ & $6p_{1/2}$ & 5.28 & 4.50 & 3.12 & 3.26 & 3.34 & 3.51 & 3.49 \\
        & $7p_{1/2}$ & 0.37 & 0.27 & -2.12 & -2.50 & -2.41 & -2.29 & -2.35 \\
        & $6p_{3/2}$ & 7.43 & 6.33 & 4.59 & 4.80 & 4.93 & 5.20 & 5.17 \\
        & $7p_{3/2}$ & 0.69 & 0.56 & -2.51 & -2.94 & -2.82 & -2.55 & -2.65 \\

        $7s_{1/2}$ & $6p_{1/2}$ & 4.41 & 4.27 & -3.76 & -4.59 & -4.44 & -4.45 & -4.48 \\
        & $7p_{1/2}$ & 11.01 & 10.30 & 6.40 & 6.91 & 6.88 & 7.19 & 7.14 \\
        & $6p_{3/2}$ & 6.67 & 6.52 & -4.75 & -5.77 & -5.56 & -5.42 & -5.49 \\
        & $7p_{3/2}$ & 15.34 & 14.31 & 9.19 & 9.90 & 9.86 & 10.29 & 10.23 \\

        $6p_{1/2}$ & $5d_{3/2}$ & 8.98 & 7.04 & -0.04 & -1.90 & -1.83 & -2.89 & -2.81 \\
        & $6d_{3/2}$ & 2.62 & 4.25 & 0.38 & 3.16 & 3.06 & 3.11 & 3.27 \\

        $6p_{3/2}$ & $5d_{3/2}$ & 4.06 & 3.17 & 0.04 & -0.83 & -0.78 & -1.33 & -1.29 \\
        & $6d_{3/2}$ & 1.34 & 2.10 & 0.39 & 1.72 & 1.67 & 1.77 & 1.83 \\
        & $5d_{5/2}$ & 12.19 & 9.69 & 0.40 & -2.23 & -2.10 & -3.32 & -3.20 \\
        & $6d_{5/2}$ & 4.02 & 6.15 & 0.54 & 4.63 & 4.48 & 4.54 & 4.70 \\

        $7p_{1/2}$ & $5d_{3/2}$ & 4.04 & 2.07 & -0.34 & -3.32 & -3.31 & -3.33 & -3.49 \\
        & $6d_{3/2}$ & 19.62 & 18.01 & 0.03 & -1.85 & -1.81 & -3.11 & -3.03 \\

        $7p_{3/2}$ & $5d_{3/2}$ & 1.69 & 0.82 & -0.30 & -1.64 & -1.64 & -1.59 & -1.68 \\
        & $6d_{3/2}$ & 8.86 & 8.08 & 0.10 & -0.83 & -0.80 & -1.53 & -1.48 \\
        & $5d_{5/2}$ & 5.02 & 2.64 & -0.31 & -4.49 & -4.49 & -4.20 & -4.45 \\
        & $6d_{5/2}$ & 26.61 & 24.38 & 0.60 & -2.18 & -2.11 & -3.76 & -3.67 \\

        \end{tabular}
    \end{ruledtabular}
\end{table}

In Table~\ref{tab:CsQED} we present QED corrections to $s-p$ and $p-d$ E1 matrix elements for Cs, showing the effect of account of many-body effects at different levels of approximation.
At each level of approximation, the QED correction is found as the difference between the result obtained with and without the radiative potential.
We present lowest-order results $|z_{ab}|$ for two approximations of particular interest: at the RHF level; and with core polarization and fitted second-order correlation corrections, which we denote as $\lambda\Sigma^{(2)}$ in the table. We present the magnitude of these values, and the signs of the QED corrections are taken to be relative to these. 
Across all transitions considered for cesium, we see that the effect of core relaxation is to increase the size of the QED correction. For the $s-p$ transitions, this increase ranges from $\sim 1\%$ to $\sim 10\%$. For the $p-d$ transitions, as we saw for francium, the account of core relaxation entirely determines the sign and size of the correction, which is produced due to QED-perturbation of the core $s$-states.
The core relaxation corrections to the $p-d$ transitions are so significant that the values reach the same order of magnitude as the QED corrections for $s-p$ matrix elements.

\begin{table*}[t]
    \footnotesize
    \caption{Reduced E1 matrix elements and QED radiative corrections for Rb, Sr$^+$, Ba$^+$, Fr, and Ra$^+$ in different approximations. The absolute values of the E1 reduced matrix elements in both the RHF and the fitted second-order correlation potential (``$\lambda\Sigma^{(2)}$'') approximations are included. The absolute QED correction in the fitted second-order correlation potential approximation is included for each atom under ``$\delta_{\text{QED}}$''. The ground-state principal quantum number $n=5$ for Rb and Sr$^+$, $n=6$ for Ba$^+$, and $n=7$ for Fr and Ra$^+$. Units: $\abs{e}a_0$.}
    \label{tab:allQED}
    \begin{ruledtabular}
        \begin{tabular}{*{17}c}
         & & \multicolumn{3}{c}{Rb} & \multicolumn{3}{c}{Sr$^+$} & \multicolumn{3}{c}{Ba$^+$} & \multicolumn{3}{c}{Fr} & \multicolumn{3}{c}{Ra$^+$} \\
        \cline{3-5} \cline{6-8} \cline{9-11} \cline{12-14} \cline{15-17}
        \multicolumn{1}{c}{$a$} & \multicolumn{1}{c}{$b$} & \multicolumn{1}{c}{RHF} & \multicolumn{1}{c}{$\lambda\Sigma^{(2)}$} & \multicolumn{1}{c}{$\delta_{\text{QED}}$} & \multicolumn{1}{c}{RHF} & \multicolumn{1}{c}{$\lambda\Sigma^{(2)}$} & \multicolumn{1}{c}{$\delta_{\text{QED}}$} & \multicolumn{1}{c}{RHF} & \multicolumn{1}{c}{$\lambda\Sigma^{(2)}$} & \multicolumn{1}{c}{$\delta_{\text{QED}}$} & \multicolumn{1}{c}{RHF} & \multicolumn{1}{c}{$\lambda\Sigma^{(2)}$} & \multicolumn{1}{c}{$\delta_{\text{QED}}$} & \multicolumn{1}{c}{RHF} & \multicolumn{1}{c}{$\lambda\Sigma^{(2)}$} & \multicolumn{1}{c}{$\delta_{\text{QED}}$} \\
        &&&&($10^{-3}$) &&& ($10^{-3}$) &&& ($10^{-3}$) &&& ($10^{-3}$) &&& ($10^{-3}$)\\

        \hline
        $ns_{1/2}$ & $np_{1/2}$ & 4.82 & 4.24 & 1.93 & 3.48 & 3.07 & 1.13 & 3.89 & 3.32 & 2.05 & 5.14 & 4.28 & 6.52 & 3.88 & 3.23 & 3.80 \\
        & $\left(n+1\right)p_{1/2}$ & 0.38 & 0.32 & -1.20 & 0.07 & 0.04 & 1.13 & 0.07 & 0.09 & 2.13 & 0.46 & 0.29 & -3.95 & 0.13 & 0.07 & 3.62 \\
        & $np_{3/2}$ & 6.80 & 5.98 & 2.77 & 4.92 & 4.34 & 1.66 & 5.48 & 4.68 & 3.14 & 7.09 & 5.89 & 10.55 & 5.34 & 4.48 & 6.57 \\
        & $\left(n+1\right)p_{3/2}$ & 0.61 & 0.52 & -1.47 & 0.16 & 0.01 & -1.39 & 0.26 & 0.04 & -2.43 & 1.10 & 0.89 & -3.24 & 0.63 & 0.36 & -3.26 \\

        $\left(n+1\right)s_{1/2}$ & $np_{1/2}$ & 4.26 & 4.16 & -2.42 & 2.38 & 2.35 & -1.76 & 2.55 & 2.51 & -3.20 & 4.53 & 4.27 & -7.49 & 2.64 & 2.55 & -5.25 \\
        & $\left(n+1\right)p_{1/2}$ & 10.29 & 9.72 & 4.05 & 6.81 & 6.53 & 2.19 & 7.39 & 7.02 & 3.87 & 10.78 & 10.08 & 13.09 & 7.37 & 6.97 & 7.08 \\
        & $np_{3/2}$ & 6.19 & 6.07 & -3.02 & 3.50 & 3.47 & -2.20 & 3.96 & 3.90 & -3.94 & 7.74 & 7.51 & -9.06 & 4.81 & 4.69 & -6.47 \\
        & $\left(n+1\right)p_{3/2}$ & 14.46 & 13.64 & 5.64 & 9.58 & 9.18 & 3.10 & 10.31 & 9.78 & 5.71 & 14.43 & 13.33 & 20.63 & 9.88 & 9.32 & 11.93 \\

        $np_{1/2}$ & $\left(n-1\right)d_{3/2}$ & 9.05 & 8.03 & -0.57 & 3.73 & 3.11 & -0.68 & 3.75 & 3.07 & -1.26 & 9.22 & 7.23 & -5.25 & 4.45 & 3.56 & -2.99 \\
        & $nd_{3/2}$ & 0.24 & 1.34 & 0.78 & 4.33 & 4.30 & 0.29 & 5.14 & 4.90 & 0.53 & 1.96 & 3.49 & 7.97 & 4.53 & 4.36 & 2.56 \\

        $np_{3/2}$ & $\left(n-1\right)d_{3/2}$ & 4.08 & 3.63 & -0.19 & 1.66 & 1.38 & -0.33 & 1.64 & 1.34 & -0.63 & 4.28 & 3.33 & -2.82 & 1.88 & 1.51 & -1.56 \\
        & $nd_{3/2}$ & 0.16 & 0.66 & 0.49 & 2.00 & 1.98 & 0.26 & 2.45 & 2.34 & 0.47 & 1.38 & 2.17 & 4.27 & 2.49 & 2.41 & 1.43 \\
        & $\left(n-1\right)d_{5/2}$ & 12.24 & 10.89 & -0.51 & 5.00 & 4.19 & -0.84 & 5.00 & 4.13 & -1.56 & 12.80 & 10.24 & -6.39 & 5.86 & 4.83 & -3.73 \\
        & $nd_{5/2}$ & 0.49 & 1.97 & 1.31 & 5.96 & 5.92 & 0.66 & 7.25 & 6.93 & 1.22 & 4.25 & 6.18 & 10.31 & 7.25 & 6.99 & 3.76 \\

        $\left(n+1\right)p_{1/2}$ & $\left(n-1\right)d_{3/2}$ & 6.73 & 5.22 & -0.99 & 0.03 & 0.05 & 0.28 & 0.35 & 0.25 & 0.40 & 4.63 & 2.57 & -8.79 & 0.11 & 0.01 & 1.70 \\
        & $nd_{3/2}$ & 18.70 & 18.20 & -0.49 & 9.09 & 8.58 & -0.85 & 9.19 & 8.67 & -1.65 & 19.83 & 18.36 & -4.71 & 10.21 & 9.57 & -4.07 \\

        $\left(n+1\right)p_{3/2}$ & $\left(n-1\right)d_{3/2}$ & 2.96 & 2.28 & -0.56 & 0.03 & 0.04 & 0.15 & 0.19 & 0.14 & 0.20 & 1.69 & 0.80 & -3.73 & 0.17 & 0.13 & 0.60 \\
        & $nd_{3/2}$ & 8.44 & 8.20 & -0.09 & 4.04 & 3.81 & -0.45 & 4.02 & 3.78 & -0.93 & 9.18 & 8.35 & -3.24 & 4.33 & 4.04 & -2.38 \\
        & $\left(n-1\right)d_{5/2}$ & 8.83 & 6.83 & -1.52 & 0.08 & 0.11 & 0.42 & 0.54 & 0.42 & 0.57 & 4.87 & 2.61 & -9.29 & 0.46 & 0.34 & 1.72 \\
        & $nd_{5/2}$ & 25.34 & 24.62 & -0.25 & 12.16 & 11.48 & -1.15 & 12.22 & 11.52 & -2.29 & 27.56 & 25.36 & -7.12 & 13.37 & 12.57 & -5.62 \\

        \end{tabular}
    \end{ruledtabular}
\end{table*}

It is seen from Table~\ref{tab:CsQED} that account of core polarization -- the electric dipole field acting on the core orbitals -- has essentially no effect on the QED corrections. Also of limited influence is the scaling of the correlation potential, mimicking the account of higher-order correlation corrections. What is important, however, is the account of correlations in the first instance. It is seen that when the correlation potential is taken into account, the QED corrections change by as much as a factor of two for some of the $p-d$ matrix elements, while the $s-p$ matrix elements hardly change. The correlation corrections for $d$ states are very large:\ for the energies, the corrections may be $\sim$\,25\% (see, e.g., Ref.~\cite{DzubaCs1983}), and for the E1 matrix elements it can be seen from  the table that in some cases they are as much as 50\%. The effect of account of the correlation potential is to pull the valence wave functions in closer to the nucleus. The likely explanation for the increase in the size of the QED corrections is the increased overlap of the $d$ states with the core $s$ states.

Therefore, we have seen that the most important many-body corrections to consider are core relaxation and account of core-valence correlations, with the former the most significant. These corrections determine the sign and magnitude of the QED corrections for $p-d$ matrix elements, while they have far less influence on the QED corrections for $s-p$ matrix elements. Still, what we see for the considered transitions in cesium is that account of many-body effects can change the QED corrections to $s-p$ matrix elements by as much as 20\%.

In Table~\ref{tab:allQED}, we present reduced E1 matrix elements and QED corrections for $s-p$ and $p-d$ transitions for the remaining atoms and ions considered in this work.
Here, we present three results for each transition: the E1 matrix elements at the RHF and ``$\lambda\Sigma^{(2)}$" approximations, and the total QED correction to the E1 matrix elements including all many-body effects that were considered for cesium in Table~\ref{tab:CsQED}.
(The QED results in Table~\ref{tab:allQED} are evaluated at the same level of approximation as those in the final column of Table~\ref{tab:CsQED}.)
As we saw for cesium, account of core relaxation changes the size of the effect on the order of 10\% for $s-p$ transitions, ranging from just a fraction of a percent, e.g., for the $8s-8p_{1/2}$ transition in Ra$^+$, through to corrections of typically about $20\%$ for $ns - (n+1)p$ and $(n+1)s - np$ transitions for all considered systems.

Indeed, for $s-p$ transitions with a change in principal quantum number, the overlap of the wave functions is reduced, and the E1 matrix elements are significantly smaller than in the case of transitions with the same $n$. However, there is no reason for suppression of the absolute QED corrections to these matrix elements. Therefore, it would be expected that the relative size of the QED corrections in these cases is enhanced. This is what we observe. We also see that the QED corrections for $s-p$ matrix elements with a change in $n$ are more sensitive to the account of many-body effects, in particular to core relaxation. This is consistent with the results for cesium.

For the $p-d$ transitions, account of core relaxation changes the magnitude and often the sign of the QED corrections, as we saw for cesium.
Without account of many-body effects, the QED corrections to $p-d$ matrix elements are typically an order of magnitude smaller than those for $s-p$ transitions (in absolute units).
Expressed as relative corrections, they are usually smaller still for $p-d$ transitions, due to the large lowest-order E1 matrix elements that is typical.
However, there are exceptions, particularly when the transition is between states with a change of 2 in principal quantum number $n$, with a correspondingly small lowest-order amplitude.
This is the case, for instance, for the $8p-6d$ transitions in Ra$^+$, where the relative corrections are as large as, or several times larger than, the corrections for the $7s-7p$ transition.

With account of many-body effects in the QED contributions, we observe that the absolute QED corrections for $s-p$ and $p-d$ transitions vary by only up to an order of magnitude across each atom or ion.
As we would expect, we see that the absolute size of the corrections increases with nuclear charge $Z$, and is smaller for the neighboring ions.
The magnitude of the lowest-order values, on the other hand, is very similar for all considered neutral atoms, and tends to be smaller for the ions.
The largest relative QED corrections, therefore, are seen generally for the heavier and more ionized systems.

\section{High-precision E1 amplitudes}
\label{sec:exptComp}

In this section, we present our results of full all-orders evaluation of E1 matrix elements, taking into account all many-body corrections described in Section~\ref{sec:theory}, and we highlight the contribution of the QED corrections to the amplitudes.
We consider those transitions for which there are high-precision experimental data available to compare with, and which have relatively significant QED corrections ($\gtrsim 0.1\%$).

\begin{table}[t]
    \caption{Final QED corrections $\delta_{\rm QED}$ and total theory values ``Theory" for E1 reduced matrix elements between states $a$ and $b$ alongside experimental values ``Expt.". Deviations $\Delta$ from experimental central values are given in the two final columns as absolute and percentage differences. Instances where $|\delta_{\rm QED}|$ is larger than $|\Delta|$ are shown in bold. Dagger ($^\dagger$) indicates that the deviation is within the experimental uncertainty. Units: $\abs{e}a_0$.}
    \label{tab:allCorrE}
    \begin{ruledtabular}
        \begin{tabular}{*{7}c}
            \multicolumn{1}{c}{$a$} & \multicolumn{1}{c}{$b$} & \multicolumn{1}{c}{$\delta_{\text{QED}}$} & \multicolumn{1}{c}{Theory} & \multicolumn{1}{c}{Expt.} & \multicolumn{1}{c}{$\Delta$} & \multicolumn{1}{c}{$\Delta$($\%$)} \\

            \hline
            \multicolumn{7}{c}{Rb} \\
            \hline
 $5s_{1/2}$    & $6p_{1/2}$ & \textbf{-0.0012} & 0.3232 & 0.3235(9)\footnotemark[1] & -0.0003 & -0.1$^\dag$ \\
            & $6p_{3/2}$ & -0.0015 & 0.5256 & 0.5230(8)\footnotemark[1] & 0.0026 & 0.5 \\

            \hline
            \multicolumn{7}{c}{Cs} \\
            \hline

            $6s_{1/2}$ & $6p_{1/2}$ & \textbf{0.0034} & 4.5052 & 4.5057(16)\footnotemark[2] & -0.0005 & -0.01$^\dag$ \\
            & $7p_{1/2}$ & \textbf{-0.0023} & 0.2776 & 0.2781(4)\footnotemark[3] & -0.0005 & -0.2 \\
            & $6p_{3/2}$ & \textbf{0.0051} & 6.3402 & 6.3398(22)\footnotemark[2] & 0.0004 & 0.01$^\dag$ \\

            & $7p_{3/2}$ & \textbf{-0.0026} & 0.5741 & 0.5742(6)\footnotemark[3] & -0.0001 & -0.01$^\dag$ \\
            $7s_{1/2}$ & $6p_{1/2}$ & -0.0044 & 4.2389 & 4.249(4)\footnotemark[4] & -0.010 & -0.2 \\

            & $6p_{3/2}$ & -0.0054 & 6.4740 & 6.489(5)\footnotemark[4] & -0.015 & -0.2 \\

            \hline
            \multicolumn{7}{c}{Ba$^{+}$} \\
            \hline

            $6s_{1/2}$ & $6p_{1/2}$ & 0.0020 & 3.3214 & 3.3251(21)\footnotemark[5] & -0.0037 & -0.1 \\
                       & $6p_{3/2}$ & 0.0031 & 4.6886 & 4.7017(27)\footnotemark[5] & -0.0131 & -0.3 \\

            \hline
            \multicolumn{7}{c}{Fr} \\
            \hline

            $7s_{1/2}$ & $7p_{1/2}$ & 0.0064 & 4.2895 & 4.277(8)\footnotemark[6] & 0.013 & 0.3 \\
            & $7p_{3/2}$ & \textbf{0.0105} & 5.9065 & 5.898(15)\footnotemark[6] & 0.009 & 0.1$^\dag$ \\

            \hline
            \multicolumn{7}{c}{Ra$^+$} \\
            \hline

            $7s_{1/2}$ & $7p_{3/2}$ & \textbf{0.0065} & 4.4827 & 4.484(13)\footnotemark[7] & -0.001 & -0.02$^\dag$ \\
            $6d_{5/2}$ & $7p_{3/2}$ & \textbf{-0.0038} & 4.7889 & 4.788(14)\footnotemark[7] & 0.001 & 0.02$^\dag$ \\

        \end{tabular}
    \end{ruledtabular}
    \footnotetext[1]{Reference \cite{Herold2012}.}
    \footnotetext[2]{Reference~\cite{Toh2019a} (average of data from  Refs.~\cite{Tanner1992,Young1994a,RafTan98,Rafac1999a,Derevianko2002a,Amini2003,Bouloufa2007,Sell2011,Zhang2013,
Patterson2015,Gregoire2015}).}
    \footnotetext[3]{Reference \cite{Damitz2019}.}
    \footnotetext[4]{Reference \cite{Toh2019}.}
    \footnotetext[5]{Reference \cite{Woods2010}.}
    \footnotetext[6]{Reference \cite{Simsarian1998}.}
    \footnotetext[7]{Reference \cite{Fan2022}.}
\end{table}

Our results are shown in Table~\ref{tab:allCorrE}. The values under the heading ``Theory" are our final results, that include the QED corrections explicitly given under the heading ``$\delta_{\rm QED}$". These theory results were found in the all-orders correlation potential method, with the final correlation potential corresponding to the fitted all-orders $\lambda\Sigma^{(\infty)}$, and with Breit correction, structural radiation, and normalization of states included. The QED values in this table are slightly different to those presented in the earlier tables, Tables~\ref{tab:CsQED} and~\ref{tab:allQED}. This is because the QED corrections have been found with account of the higher-order correlation corrections; see Eq.~(\ref{eq:QEDcorr}). We refer the reader to Section~\ref{sec:theory} for more information about the approach; a more detailed description of the method, along with results spanning many more transitions, can be found in Ref.~\cite{Roberts2022E1}.

In the remaining three columns of Table~\ref{tab:allCorrE}, we present the experimentally-determined E1 matrix elements and the deviations between our results and experiment (in absolute units and as a percentage). For all considered cases, the deviations are within several $0.1\%$. And in fact, for most cases, the deviation is only 0.1\% or less. This level of agreement for high-precision many-body calculations for heavy atoms and ions is unprecedented. The account of QED corrections, while small, makes a meaningful contribution to the final many-body results in several instances. Those cases for which the size of the QED correction exceeds the deviation of the final theory result from experiment are indicated in boldface. Notably, this includes four transitions in cesium, $6s-6p$ and $6s-7p$; it is worth mentioning that these transitions are particularly important for the calculation of the cesium atomic parity violating amplitude~\cite{GingesCs2002,Porsev2009,*Porsev2010,DzubaCsPNC2012}. This demonstrates that high-precision many-body atomic calculations have reached the level of precision where QED contributions are important.

We note that while the deviations of the theory values from experiment give an indication of the accuracy of atomic many-body methods, it is also possible to assign reliable estimates for the uncertainty of the theory calculations from first principles.
This is studied at great length in our paper~\cite{Roberts2022E1}, where a detailed survey of theoretical and experimental E1 matrix elements is carried out, and a robust estimate for the theory uncertainty for the all-orders correlation potential method is proposed and tested.

\section{Conclusion}

We have used the radiative potential method to carry out a detailed study of the combination of QED and many-body effects for electric dipole matrix elements.
From comparison with rigorous QED, we have found that the radiative potential method can be used to evaluate corrections to E1 matrix elements for heavy atoms with an accuracy of about $25\%$ or better for most atoms.
Our study of the influence of many-body effects shows that account of core relaxation and valence-core electron correlations are particularly important, and can change the QED correction by an order of magnitude.
For transitions with a change in principal quantum number, the relative QED corrections may be large, and they could be exploited to test QED in heavy atoms.
We have also performed all-orders many-body calculations for full E1 matrix elements.
Comparison with experiment shows that the calculations of E1 matrix elements have reached the level of about $0.1\%$ uncertainty.
There are several transitions for which the QED contribution is larger than the deviation of the calculated E1 matrix element from experiment.
That is, atomic theory has reached the level of precision where the QED corrections to E1 amplitudes may be resolved in heavy-atom experiments.

\acknowledgements

We thank J.\ Brown and B.\ Oxley for their involvement in the early stages of this work.
This work was supported by the Australian Government through an Australian Research Council Future Fellowship No.~FT170100452 and DECRA Fellowship No.\ DE210101026.

\bibliography{qed-e1}

\end{document}